\documentclass[12pt]{article}
\usepackage[compress]{cite}
\usepackage{graphicx}
\usepackage{epsfig}
\usepackage{amssymb}
\begin{document}

\begin{center}
{\bf
Unified description of LHC data on elastic $pp$ scattering
}

\vspace{.2cm}

Yu.M. Shabelski and A.G. Shuvaev \\

\vspace{.5cm}

Petersburg Nuclear Physics Institute, Kurchatov National
Research Center\\
Gatchina, St. Petersburg 188300, Russia\\
\vskip 0.9 truecm
E-mail: shabelsk@thd.pnpi.spb.ru\\
E-mail: shuvaev@thd.pnpi.spb.ru

\vspace{1.2cm}

\end{center}

\begin{abstract}
\noindent
We present the unified description of the existing
data
on elastic small angle $pp$ and $p \bar p$ scattering at
the energies
1.8, 1.96, 2.76, 7, 8 and
13~TeV in the framework of Additive Quark Model.
The agreement with the data is quite reasonable.

\end{abstract}

\section{Introduction}

In the previous papers \cite{Shabelski:2014yba,
Shabelski:2015bba,Shabelski:2016aek}
we describe the differential cross
section of elastic $pp$ scattering at $\sqrt s =7$~TeV.
The real part of $pp$
elastic amplitude measured at $\sqrt s =13$~TeV has been described
in \cite{Shabelski:2018jfq}.
Recently the new data on elastic scattering cross section
$d\sigma/dt$ became available
at the energies 2.76, 8 and 13~TeV. In the present paper we
show that these data can be reasonable described
in the same manner
after a little modification of the internal parameters.

Our approach is based on the Additive Quark Model (AQM).
In AQM baryon is treated as a system of
three spatially separated compact objects -- the constituent quarks.
Each constituent quark is colored and has an internal quark-gluon
structure and a finite radius that is much less than the radius of
the proton, $r_q^2 \ll r_p^2$.
The constituent quarks play the roles of incident particles
in terms of which $pp$ scattering is described.
Elastic amplitudes for large energy $s=(p_1+p_2)^2$
and small momentum transfer $t$ are dominated by Pomeron
exchange. We neglect the small difference in $pp$ and $p\bar p$
scattering coming from the exchange of negative signature
Reggeons such as
the Odderon,
$\omega$-Reggeon etc. (see e.g.~\cite{Lukaszuk:1973nt, Avila,
Martynov:2018pye, Fried:2019afs}, since their contributions
are suppressed by $s$.
The single $t$-channel exchange results into
the amplitude of constituent quarks scattering,
\begin{equation}
\label{Mqq}
A_{qq}^{(1)}(s,t) = \gamma_{qq}(t) \cdot
\left(\frac{s}{s_0}\right)^{\alpha_P(t) - 1} \cdot
\eta_P(t) \;,
\end{equation}
where $\alpha_P(t) = \alpha_P(0) + \alpha^\prime_P\cdot t$
is the Pomeron trajectory specified by the intercept
$\alpha_P(0)$ and the slope value $\alpha^\prime_P$.
The Pomeron signature factor,
$$
\eta_P(t) \,=\, i \,-\, \tan^{-1}
\left(\frac{\pi \alpha_P(t)}2\right),
$$
determines the complex structure of the amplitude. The factor
$\gamma_{qq}(t)=g_1(t)\cdot g_2(t)$ has the meaning
of the Pomeron coupling to the beam and target particles,
the functions $g_{1,2}(t)$ being the vertices of the constituent
quark-Pomeron interaction.
In the following we assume the Pomeron trajectory
to have the simplest form,
$$
\left(\frac{s}{s_0}\right)^{\alpha_P(t) - 1}\,=\,e^{\Delta\cdot\xi}
e^{-r_q^2\,q^2}, ~~ \xi\equiv \ln\frac{s}{s_0},~~
r_q^2\equiv \alpha^\prime\cdot\xi.
$$
The value $r_q$ defines the radius of the quark-quark interaction,
while $S_0=(9~{\rm GeV})^2$ has the meaning of typical energy scale
in Regge theory.
The scattering amplitude is presented in AQM as a sum over
the terms with a given number of Pomerons,
\begin{equation}
\label{totamp}
A_{pp}(s,t)\,=\,\sum_n A_{pp}^{(n)}(s,t),
\end{equation}
where the amplitudes $A_{pp}^{(n)}$ collect all
the diagrams comprising various interactions
between the beam and target quarks going through
$n$ Pomerons exchange.
Similar to Glauber theory \cite{Glaub, FG}
the multiple interactions between the same
quark pair has to be ruled out.
AQM permits the Pomeron to connect any two quark only once.
It crucially decreases the combinatorics, leaving the diagrams
with no more than $n=9$ effective Pomerons.
If $q_i$ are the momenta carrying by a given Pomeron,
then
\begin{eqnarray}
\label{Mn}
A_{pp}^{(n)}(s,t)\,&=&\,i^{n-1}\biggl(\gamma_{qq}\eta_P(t_n)
e^{\Delta\cdot\xi}\biggr)^n\,
\int\frac{d^2q_1}{\pi}\cdots \frac{d^2q_n}{\pi}
\,\pi\,\delta^{(2)}(q_1+\ldots +\,q_n-Q)\,\\
&&\times\,e^{-r_q^2(q_1^2+\ldots + q_n^2)}\,
\frac 1{n!}\sum\limits_{n~\rm connections}\hspace{-1.5em}
F_P(Q_1,Q_2,Q_3)\,F_P(Q_1^{\,\prime},Q_2^{\,\prime},Q_3^{\,\prime}),
~~~t_n\simeq t/n. \nonumber
\end{eqnarray}
The sum in this formula refers to all distinct ways
to connect the beam and target quark lines with $n$ Pomerons
in the scattering diagram. The set
$Q_i$ and $Q_l^{\,\prime}$ stands for the momenta
the quarks acquire from the Pomerons attached to it.
It is particular for each connection pattern.
$Q$ is the total momentum transferred in the scattering,
$t=-Q^2$.
The Pomeron-proton form factor,
\begin{equation}
\label{FP}
F_P(Q_1,Q_2,Q_3)\,=\,\int dk_i\,\psi^*(k_1,k_2,k_3)\,
\psi(k_1+Q_1,k_2+Q_2,k_3+Q_3),
\end{equation}
is expressed through the initial proton
wave function $\psi(k_1,k_2,k_3)$
written in terms of the constituent quarks'
transverse momenta $k_i$, and
the wavefunction of the scattered proton,
$\psi(k_1+Q_1,k_2+Q_2,k_3+Q_3)$.
A more detailed description can be found
in~Ref.~\cite{Shabelski:2014yba}.

The quarks' wave function has been taken in the simple form
of Gaussian packets,
\begin{equation}
\label{gausspack}
\psi(k_1,k_2,k_3)\,=\,N\bigl[\,e^{-a_1(k_1^2+k_2^2+k_3^2)}\,
+\,C_1\,e^{-a_2(k_1^2+k_2^2+k_3^2)}
+\,C_2\,e^{-a_3(k_1^2+k_2^2+k_3^2)}\bigr],
\end{equation}
normalized to unity,
$$
\int \bigl|\psi(k_1,k_2,k_3)\bigr|^2\delta^{(2)}(k_1+k_2+k_3)\,
d^2k_1 d^2k_2 d^2k_3\,=\,1.
$$

\section{Comparison with the data}

The results of our calculations are presented
for the relatively small $t$. The large $t$ values
are beyond the validity of AQM since it treats
the constituent quarks as point like particles
without an internal structure.
Differential cross section $d\sigma/dt$
are shown in Fig.1~(left)
together with the experimental data
for $\sqrt s = 13$~TeV. The model provides fairly
well fit to the data for $|t| <0.6$~GeV$^2$.
The results for $\sqrt s = 8$~TeV are shown
in Fig.1~(right).
Unfortunately the experimental data
are available at this energy
not up to the dip position.

\begin{figure}[htb]
\vskip -1.cm
\includegraphics[width=.45\textwidth]{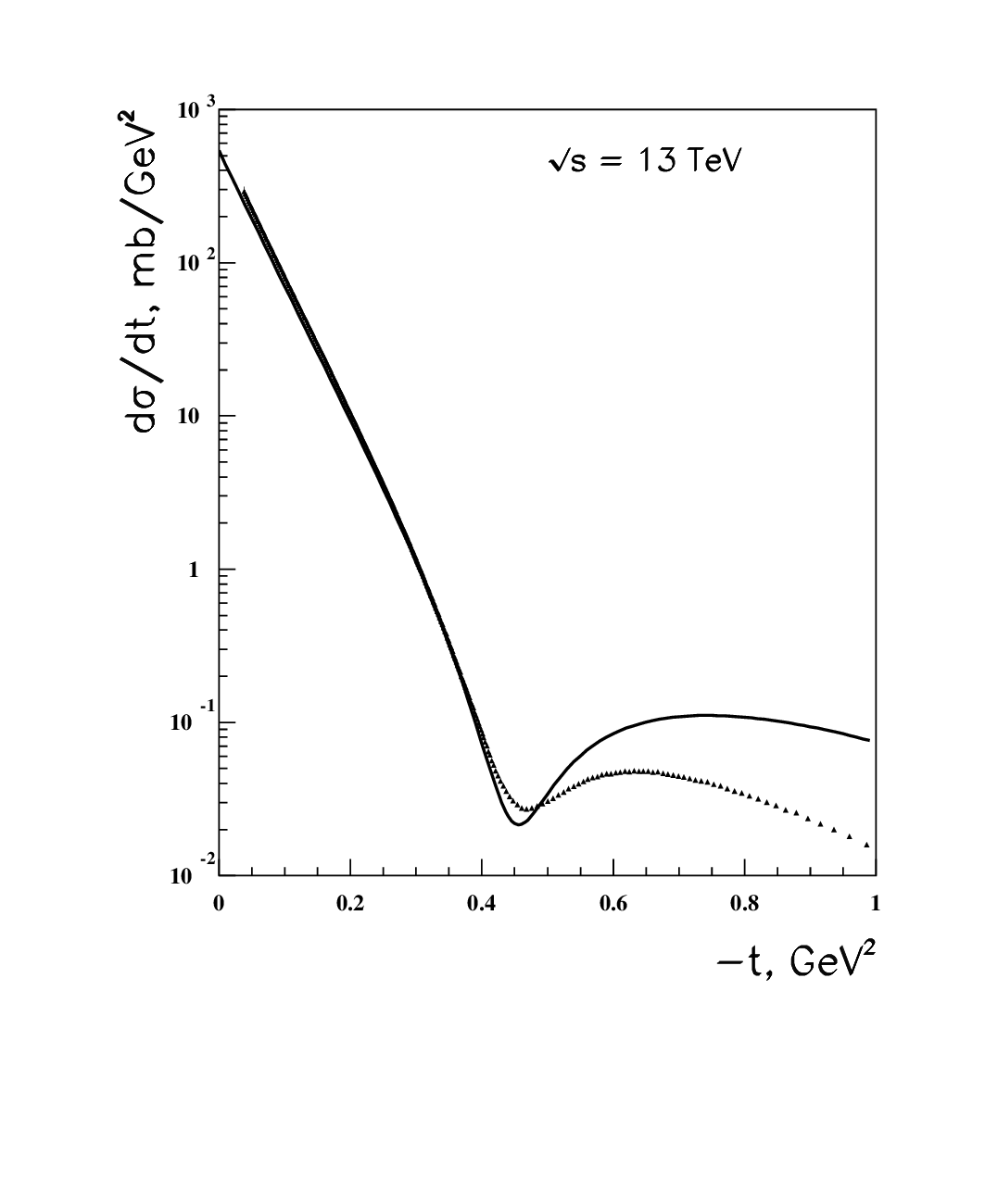}
\includegraphics[width=.45\textwidth]{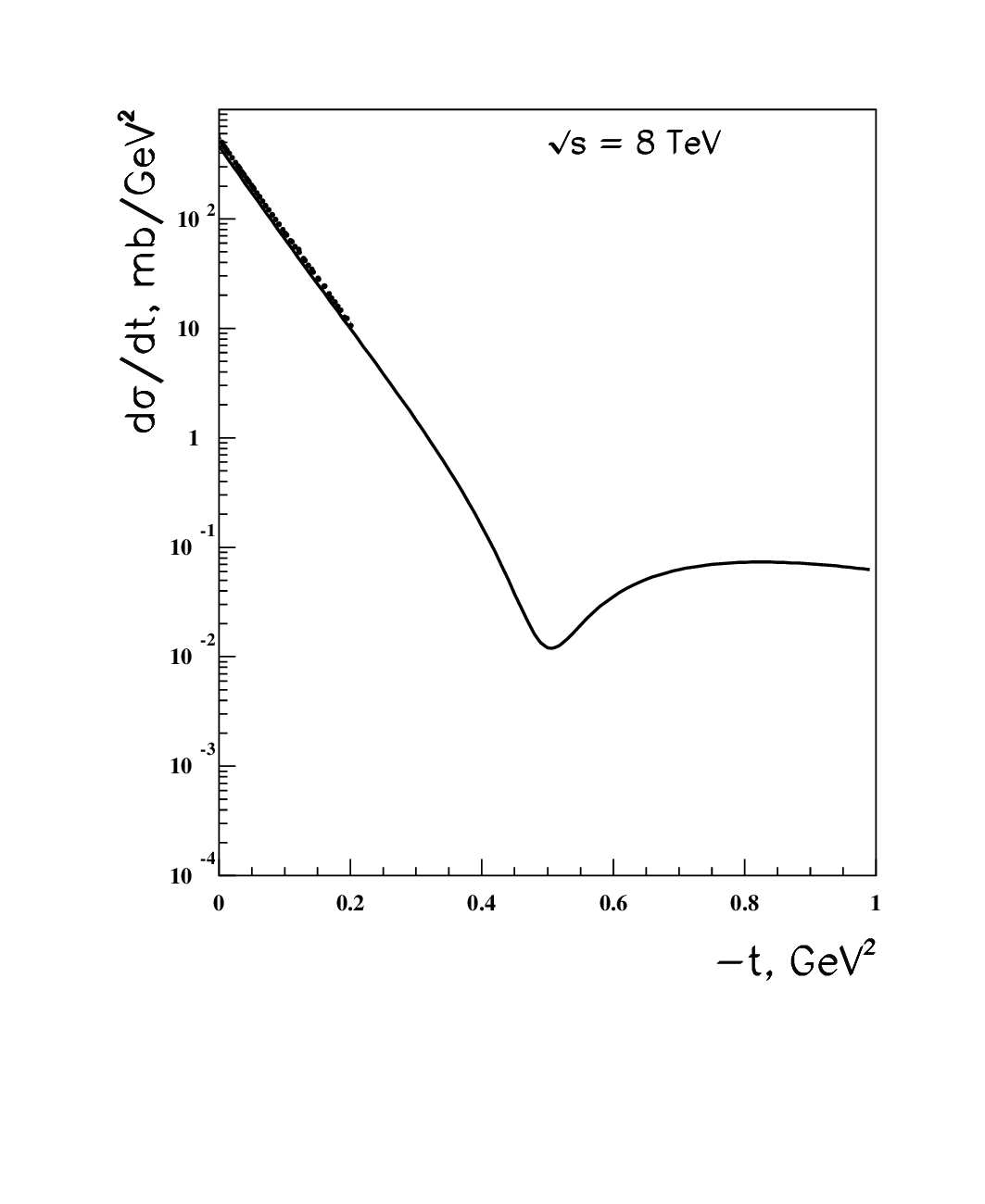}
\vskip -1.5 cm
\caption{\footnotesize
The differential cross section of elastic $pp$ scattering
at $\sqrt s = 13$~TeV (left) and $\sqrt s = 8$~TeV
(right).
The experimental points have been taken from
\cite{Antchev:2018edk,Antchev:2017dia,
Antchev:2015zza,Antchev:2016vpy}.}
\end{figure}
\begin{figure}[htb]
\vskip -1.cm
\includegraphics[width=.45\textwidth]{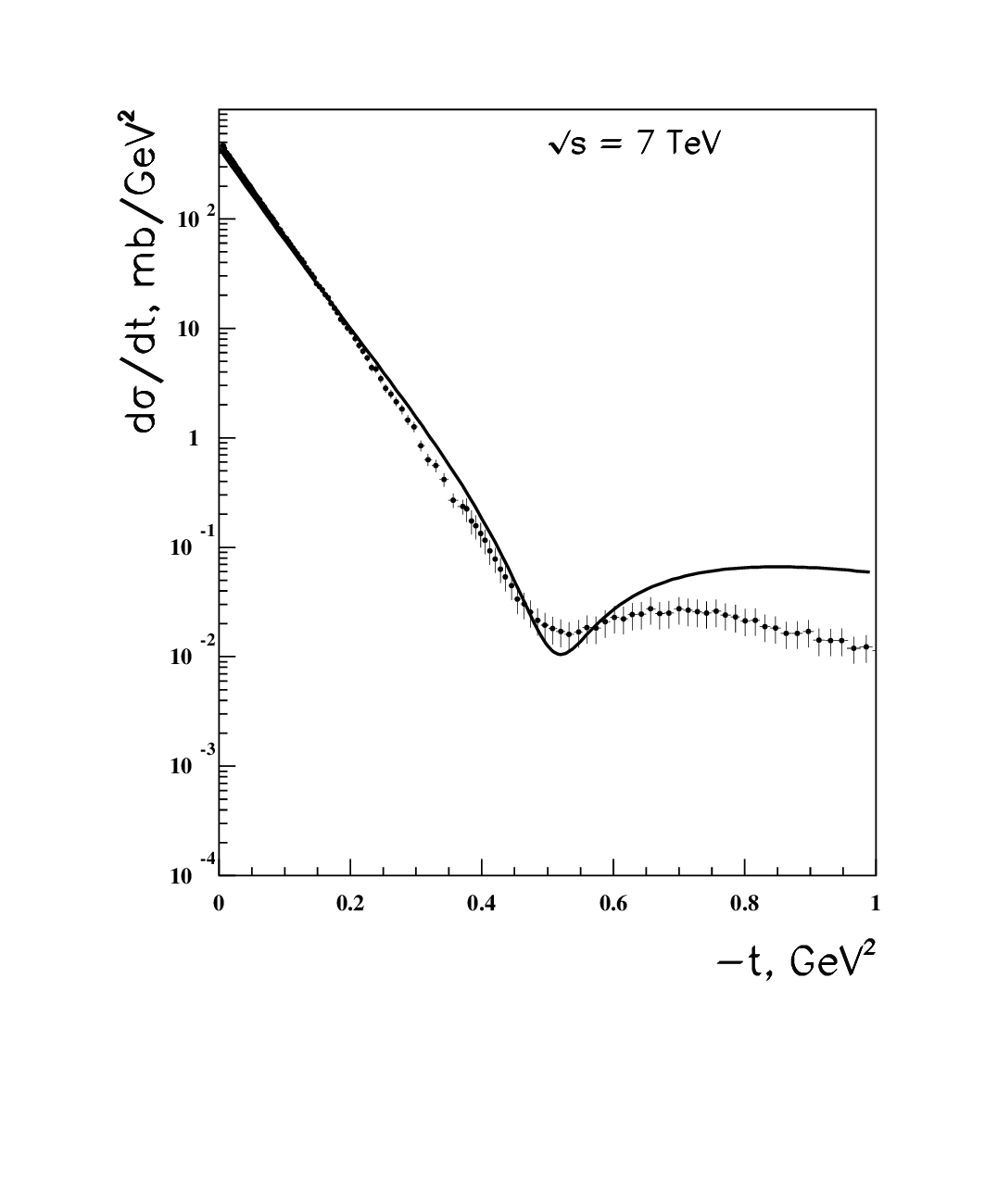}
\includegraphics[width=.45\textwidth]{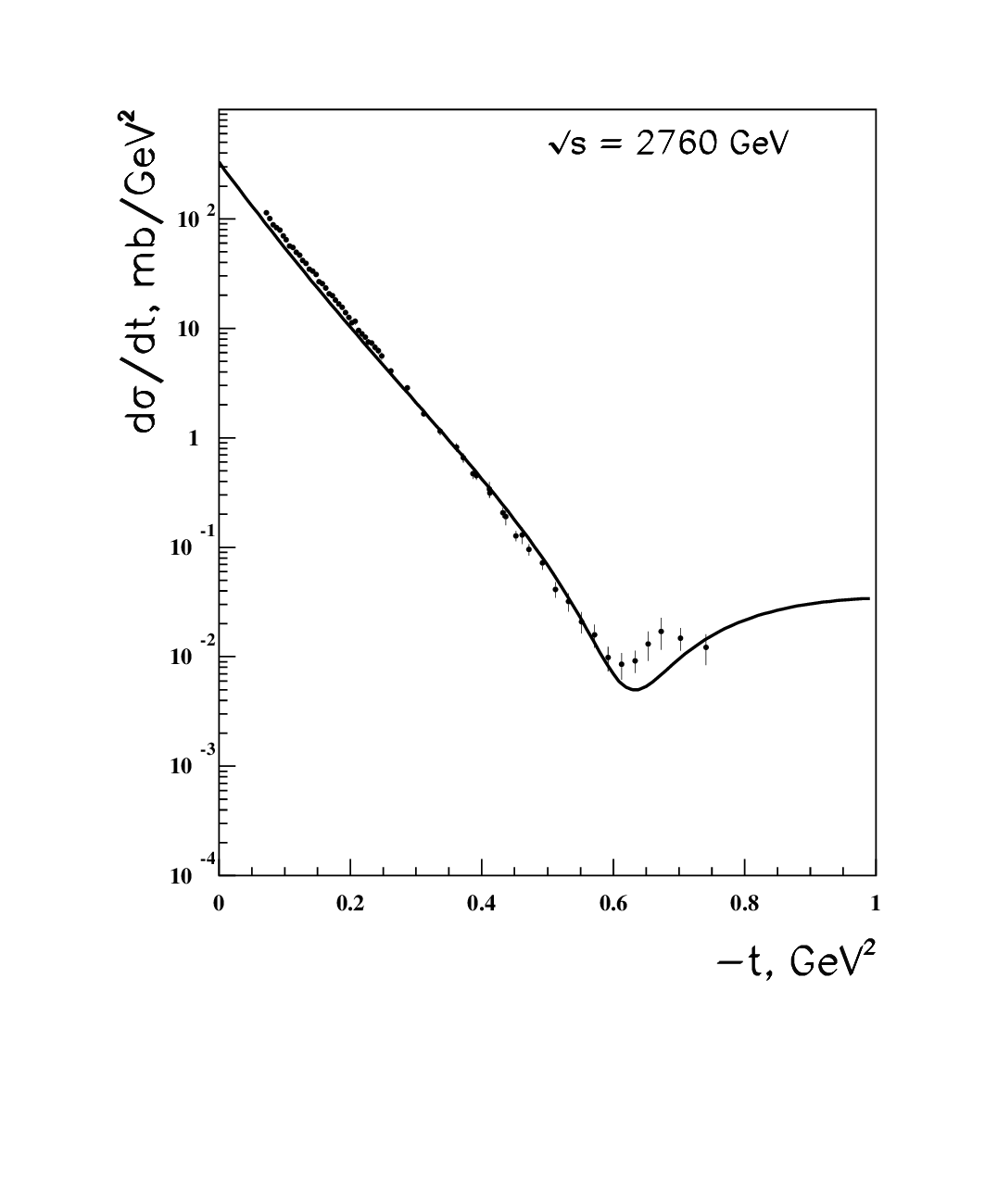}
\vskip -1.5 cm
\caption{\footnotesize
The differential cross section of elastic $pp$ scattering
at $\sqrt s = 7$~TeV (left) and $\sqrt s = 2.76$~TeV
(right).
The experimental points have been taken from
\cite{TO2, TO1, Antchev:2018rec}
.}
\end{figure}

The calculated ratio of the real to imaginary
part of the elastic scattering amplitude at $t=0$
is $\rho = 0.82$ for $\sqrt s = 13$~TeV.
It agrees with the experimental value
$0,9 \pm 0.01$~\cite{Antchev:2017yns}
while being only a little bit smaller
than the other value, $0.10 \pm 0.01$,
reported in the same Ref.~\cite{Antchev:2017yns}.

Fig.2 presents $d\sigma /dt$ data
for $\sqrt s=7$~TeV and $\sqrt s=2.76$~TeV.
The calculations for 7~TeV yield an appropriate
fit to the data though the output total cross section
$\sigma_{tot} = 93.1$~mb is smaller than the experimental
estimation $98.6 \pm 2.2$~mb~\cite{TO1}.
The position of the minimum obtained at 2.76~TeV
is a little more to the right than the experimental one.
It is important to note that the values of the total
cross section, slope of the diffractive cone and
the location of the minima of the differential
cross section are strongly correlated at the each
energy.

\begin{figure}[htb]
\vskip -1.cm
\includegraphics[width=.45\textwidth]{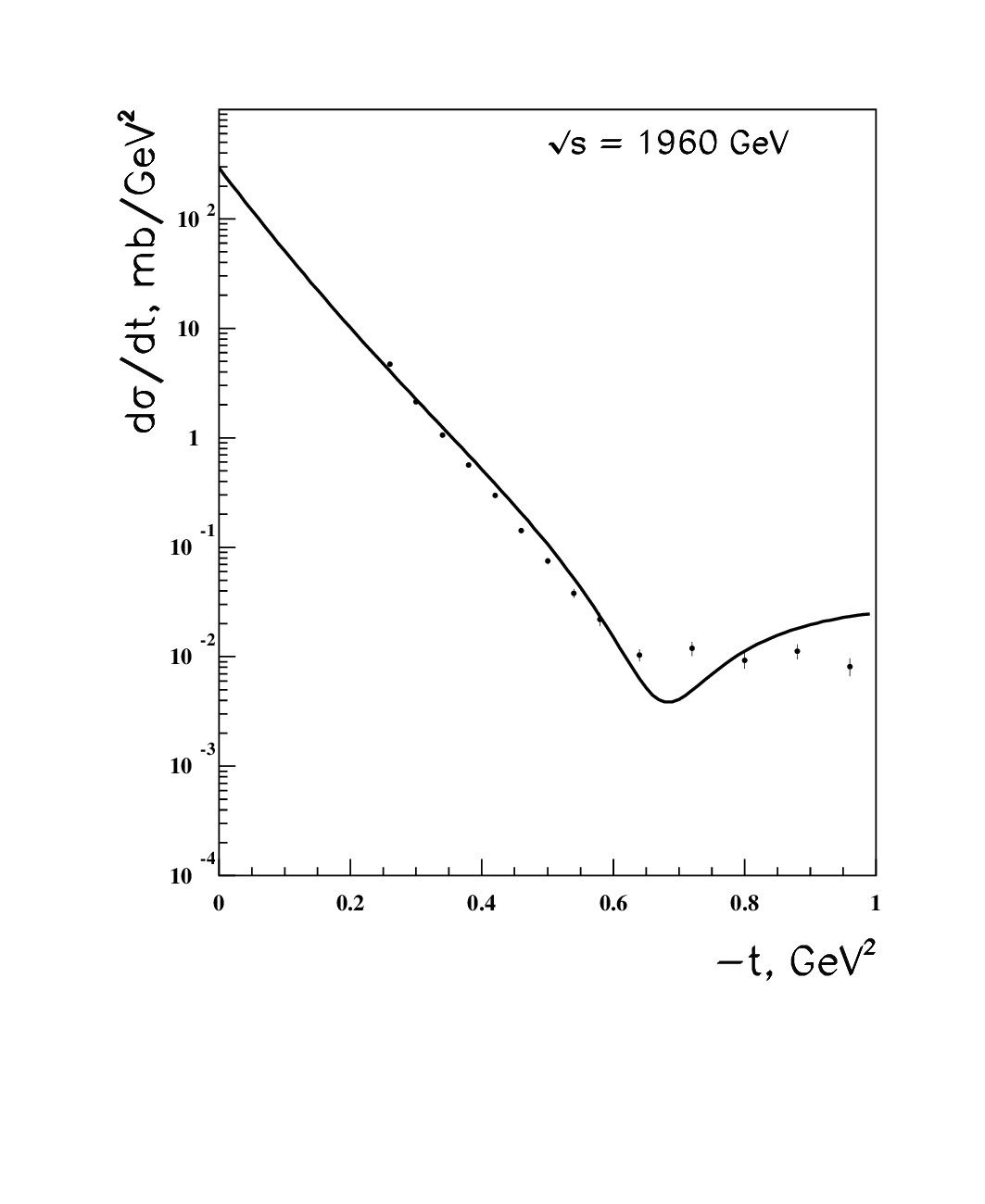}
\includegraphics[width=.45\textwidth]{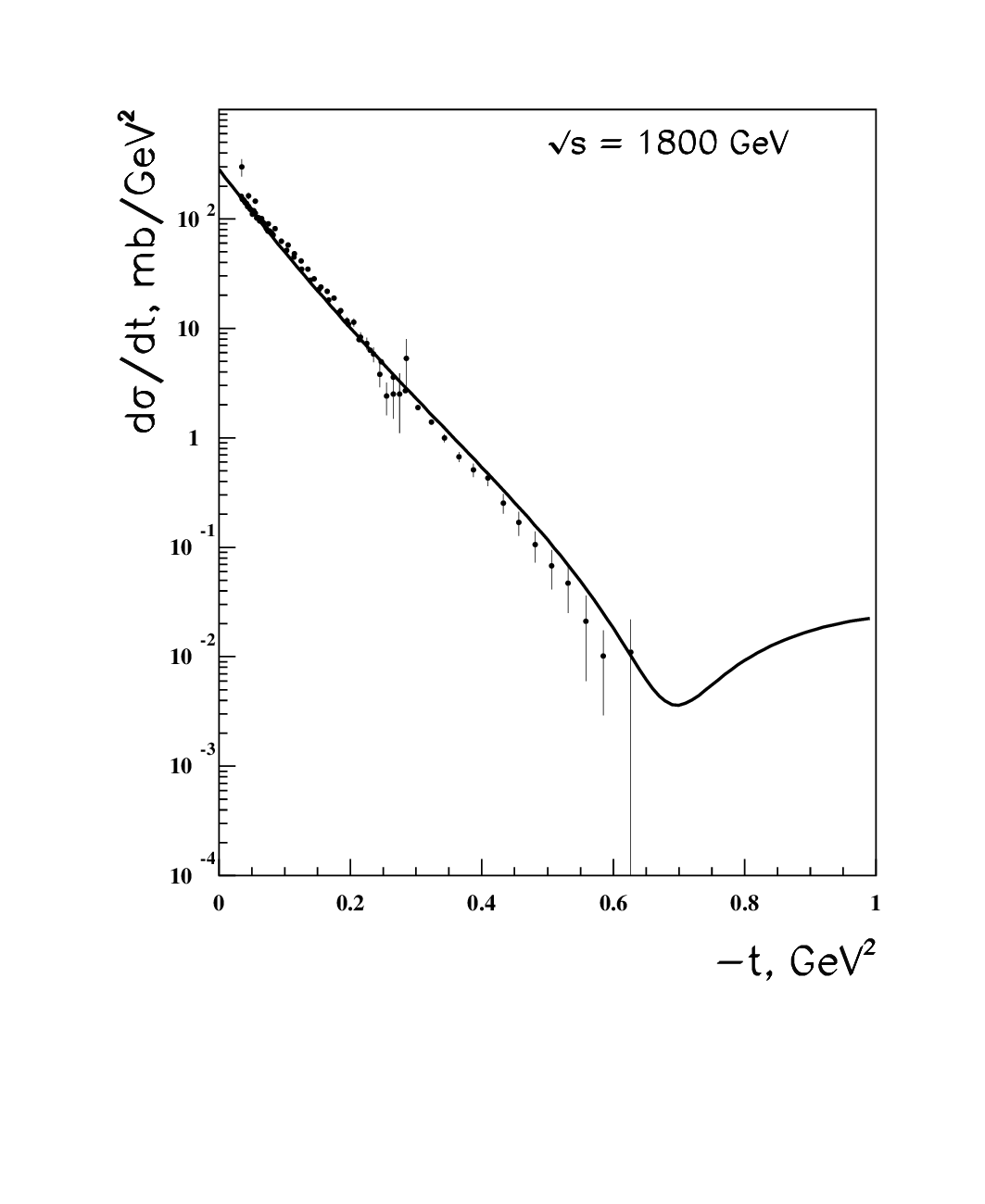}
\vskip -1.5 cm
\caption{\footnotesize
The differential cross section of elastic $p\bar p$ scattering
at $\sqrt s = 1.96$~TeV (left) and $\sqrt s = 1.8$~TeV (right).
The experimental points have been taken from
\cite{Abazov:2012qb, Abe:1993xx, Amos:1990fw,
Amos:1991bp}.}
\end{figure}

To better illustrate applicability of the model
we also present what it gives for the $p\bar p$ elastic
scattering at the lower energies. The comparison
with the FNAL data at the energies 1.96~Tev
and 1.8~TeV, Fig.3, exhibits a quite acceptable
description, however the dip at 1.96~TeV is again
right of its experimental position.

All model curves have been obtained with the following
set of parameters:
\begin{equation}
\label{param}
\Delta=0.139,~~~
\alpha^\prime=0.122\,{\rm GeV}^{-2},~~~
\gamma_{qq}=0.49\,{\rm GeV}^{-2}.
\end{equation}
$$
a_1=8.0\,{\rm GeV}^{-2},~~~
a_2=0.255\,{\rm GeV}^{-2},~~~
a_3=1.45\,{\rm GeV}^{-2},~~~
C_1=0.026,~~~
C_2=0.054.
$$

The calculated total cross section, $\sigma_{tot}$,
slope of the diffractive
cone $B$ ($d\sigma/dt \propto\,\exp(-B]t|)$) and the ratio
of the real to imaginary part of the elastic $pp$
(or $\bar pp$) amplitude,
$\rho\,=\,\mathrm{Re}\,A/\mathrm{Im}\,A$,
are compared with the existing
experimental data in the Table.

\begin{center}
\begin{tabular}{|c||c|c|c|}
\multicolumn{4}{c}{\parbox{.8\textwidth}{\footnotesize
Comparison of the calculated total cross section,
$\sigma_{tot}$, slope of the diffractive
cone,$B$, and  the ratio,
$\mathrm{Re}\,A/\mathrm{Im}\,A$,
of the real to the imaginary part
of the elastic
$pp$ (or $\bar pp$) amplitude with the available
experimental data.

\medskip}
}\\
\hline
$\sqrt{s}$  & $\sigma_{tot}$ (mb) &
$B$ (GeV$^{-2}$) & ${\rm Re}\,A/{\rm Im}\,A$ \\
& & (mb/GeV$^2$)& $(t=0)$ \\ \hline
13 TeV & 102.4 & 20.4 & 0.082 \\
\cite{Antchev:2017dia}& $110.5 \pm 2.4$& & \\
\cite{Antchev:2018edk} & & $20.40 \pm 0.002 \pm 0.01$& \\
\cite{Antchev:2017yns} & & & $0.09 \pm 0.01$ \\
\cite{Antchev:2017yns} & & & $0.10 \pm 0.01$ \\ \hline
8 TeV & 95.1 & 19.6 & 0.083 \\
\cite{Antchev:2015zza}& $101.5 \pm 2.1$ & & \\
\cite{Antchev:2015zza}& $101.9 \pm 2.1$ & & \\
\cite{Antchev:2016vpy}& $102.0 \pm 2.3$ & & $0.12 \pm 0.03$\\
\cite{Antchev:2016vpy}& $103.0 \pm 2.3$ & &  \\ \hline
7 TeV & 93.1 & 19.4 & 0.083 \\
\cite{TO2} &  & $23.6 \pm 0.5 \pm 0.4$ & \\
\cite{TO1} & $98.6 \pm 2.2$ & $19.9 \pm 0.3$ & \\ \hline
2.76 TeV & 80.0 & 18.0 & 0.087 \\
\cite{Antchev:2018rec}& & $17.1 \pm 0.3$ & \\ \hline
1.96 TeV & 75.5 & 17.5 & 0.089 \\
\cite{Abazov:2012qb}& & $16.86 \pm 0.1 \pm 0.2$ & \\ \hline
1.8 TeV & 74.4 & 17.4 & 0.089 \\
\cite{Amos:1990fw} & & $16.99 \pm 0.47$ & \\
\cite{Amos:1991bp} & & & $0.14 \pm 0.069$ \\ \hline
\end{tabular}
\end{center}
\noindent
\vskip-0.2cm

\section{Discussion and conclusion}

As is seen from the Table the calculated cross sections
are smaller than the experimental ones at the LHC energies.
The output $\sigma_{tot}$ values can be increased
by readjusting the model parameters, but it moves
the diffractive dip to the left and causes the growth
of the $\rho$ ratio. Given the fact that the experimental
dip position is directly measured whereas
an additional analysis is needed to get
the absolute normalization of the cross section
(see e.g.~\cite{Selyugin:2019ybv}),
we prefer to chose the above
listed parameters~(\ref{param}).
Despite some mismatch in the experimental
slopes $B$ of the diffractive cones at small $t$
the general trend of its
energy dependence is properly reproduced.

To summarize, the conventional AQM with a rather simple
set of parameters results in quite reasonable description
of the small angle hard scattering. Though we are unable
to exclude completely effects from the Odderon
exchange~\cite{Antchev:2018rec} their contribution
seems not to be significant. Here we agree with
Ref.~\cite{Khoze:2018kna}.

The authors are grateful to M.G. Ryskin for helpful discussion.

\end{document}